\documentstyle[12pt]{article}
\begin{document}
\noindent
{\large\bf UN/ESA Workshops on Basic Space Science: An Update on Their Achievements}
\begin{center}
{\bf Hans J. Haubold}\\
Programme on Space Applications\\
Office for Outer Space Affairs\\
United Nations\\
Vienna International Centre\\
P.O. Box 500\\
A-1400 Vienna, Austria\\
Email: haubold@kph.tuwien.ac.at
\end{center}\par
\bigskip
\noindent
Abstract\par
\smallskip
\noindent
During the second half of the twentieth century, expensive observatories are being erected at La Silla (Chile), Mauna Kea (Hawai), Las Palmas (Canary Island), and Calar Alto (Spain), to name a view. In 1990, at the beginning of The Decade of Discovery in Astronomy and Astrophysics (Bahcall [2]), the UN/ESA Workshops on Basic Space Science initiated the establishment of small astronomical telescope facilities, among them many particularly supported by Japan, in developing countries in Asia and the Pacific (Sri Lanka, Philippines), Latin America and the Caribbean (Colombia, Costa Rica, Honduras, Paraguay), and Western Asia (Egypt, Jordan, Morocco). The annual UN/ESA Workshops continue to pursue an agenda to network these small observatory facilities through similar research and education programmes and at the same time encourage the incorporation of cultural elements predominant in the respective cultures. Cross-cultural integration and multi-lingual scientific cooperation may well be a dominant theme in the new millennium (Pyenson [20]). This trend is supported by the notion that astronomy has deep roots in virtually every human culture, that it helps to understand humanity's place in the vast scale of the Universe, and that it increases the knowledge of humanity about its origins and evolution. Two of these Workshops have been organized in Europe (Germany 1996 and France 2000) to strengthen cooperation between developing and industrialized countries.\par
\bigskip
\noindent
{\bf 1. Introduction}\par
\medskip
\noindent
Answering questions about the Universe challenges astronomers, fascinates a broad national audience and inspires young people to pursue careers in engineering, mathematics, and science. Basic space science research assists nations, directly and indirectly, in achieving societal goals. For example, studies of the Sun, the planets, and the stars have led to experimental techniques for the investigation of the Earth's environment and to a broader perspective from which to consider terrestrial environmental concerns such as
ozone depletion and the greenhouse effect [1]. \par
\smallskip
\noindent
Basic space science makes humanistic, educational and technical contributions to society. The most fundamental contribution of basic space science is that it provides modern answers to questions about humanity's place in the Universe. Quantitative answers can now be found to questions about which ancient philosophers could only speculate. In addition to satisfying curiosity about the Universe, basic space science nourishes a scientific outlook in society at large. Society invests in basic space science research and receives an important dividend in the form of education, both formally through instruction in schools, colleges and universities, and more informally through television programmes,
popular books and magazines, and planetarium presentations. Basic space science introduces young people to quantitative reasoning and also contributes to areas of more immediate practicality, including industry, medicine and the understanding of the Earth's environment [2].\par
\smallskip
\noindent
The international basic space science community has long shown leadership in initiating international collaboration and cooperation. Forums have been established on a regular basis in which the basic space science community has publicized its scientific achievements and the international character of astronomical and space science studies. The most recent such initiatives were the International Space Year (1992), with its elements Mission to Planet Earth and Mission to the Universe, and the Third United Nations Conference on the Exploration and Peaceful Uses of Outer Space (UNISPACE III), held from 19-30 July 1999 at the United Nations Office Vienna, Austria [21].\par
\smallskip
\noindent 
Despite the considerable progress made in the development of astronomy and basic space science, the observation has been made that of the 188 countries that are Member States of the United Nations, nearly 100 have professional or amateur astronomical organizations. Only about 60 of these countries, however, are sufficiently involved in astronomy to belong to the International Astronomical Union. Only about 20 countries, representing 15\% of the world's population, have access to the full range of astronomical facilities and information. This does not include most of the Eastern European, Baltic, and former countries of the Soviet Union, whose fragile economies keep them from achieving their full potential, despite the excellence of their astronomical heritage and education [3].\par 
\medskip
\noindent
{\bf 2. First Cycle of Workshops: Regional Observations and Recommendations}\par
\smallskip
\noindent
In 1991, the United Nations, through its Programme on Space Applications of the UN Office for Outer Space Affairs, in cooperation with the European Space Agency, held its first Workshop on Basic Space Science in India for Asia and the Pacific region [4]. Since then, such workshops have been held annually in the different regions around the world to make a unique contribution to the worldwide development of astronomy and space science, particularly in developing countries. Workshops were held in 1992 in Costa Rica and Colombia for Latin America and the Caribbean, in 1993 in Nigeria for Africa, and in 1994 in Egypt for Western Asia. Additionally to the direct benefits of a common research workshop, a vital part of each of the workshops were daily working group sessions, which provided participants a forum in which observations and recommendations for developing basic space science in all its aspects, through regional and international cooperation, have been made. The deliberations of these sessions and the observations and recommendations that emanated from them, region by region, are published as UN General Assembly documents that can be used to lobby governments and funding agencies to implement prospective follow-up projects of the workshops [5].\par
\medskip
\noindent 
{\bf 3. Second Cycle of Workshops: Implementing Follow-up Projects}\par
\smallskip
\noindent
Among the most important results emanating from the workshop series, starting with the first workshop in India in 1991, is that for each of the regions a number of follow-up projects were identified, mainly the establishment and operation of small astronomical telescope facilities, which have been gradually implemented over the course of the workshops. Examples of them are briefly listed below.\par
\smallskip
\noindent
The Galactic Emission Mapping (GEM) project of researchers from Brazil, Colombia, Italy, Spain, and the United States, devised to obtain a full sky, multi-frequency, and high sensitivity survey of the galactic radio emission, lead to the operation of the GEM radio telescope at an equatorial site in Colombia in 1995. Subsequently, in order to cover the northern and southern latitudes not visible from this equatorial site, the radio telescope was moved to Spain (IAC at Tenerife) and to Brazil (INPE at Sao Jose dos Campos) to continue radio frequency observations. Since 1995, the results obtained with GEM are\par
\smallskip
\noindent
(i) a galactic radio emission database in the frequencies 408, 2300, and 5000 MHz;\\
(ii) an estimate of the sky temperature and spectral indices within that frequency range; and\\ 
(iii) an estimate of the galactic emission profile and quadrupole component [6].\par
\smallskip
\noindent
In 1995, the workshop was held in Sri Lanka to inaugurate an astronomical telescope facility, based on a donation of an astronomical telescope (45-cm GOTO) from Japan to Sri Lanka, at the Arthur C. Clarke Institute for Modern Technologies. The telescope is equipped with a photoelectric photometer, spectrograph, and an ordinary camera for imaging (recently, a CCD camera was installed), and necessary computer equipment. Young astronomers are being trained and educated for the operation of the telescope facility through comprehensive programmes at Bisei Observatory, Japan. Since 1995, the ACCIMT serves as the national centre for research, education, and popularization
of astronomy in Sri Lanka [7].\par
\smallskip
\noindent
In 1997, the workshop inaugurated the Central American Astronomical Observatory (utilizing a 40-cm Maede telescope) at the National Autonomous University of Honduras, Tegucigalpa, Honduras, with the dedication of the Telescopio Rene Sagastume Castillo at the Suyapa Observatory for Central American countries (Costa Rica, El Salvador, Guatemala, Honduras, Nicaragua, Panama) [8].\par
\noindent
Following the recommendation of the workshop in Egypt in 1994, the long awaited
refurbishment and modernization of the 74" Kottamia telescope at Helwan, Egypt, will be finalized in 1999. This telescope, equipped  with Cassegrain and Coude spectrographs, saw first light in 1964 and is still the  largest telescope in the region of Western Asia and will be made available for regional and international cooperation in the near future. The agreement between the National Research Institute of Astronomy and Geophysics (NRIAG) and the Government of Egypt lead to the replacement of the primary and secondary mirrors of the telescope by new mirrors of Zerodur glass ceramics. To improve the optical performance of the telescope system, a more efficient supporting system was also developed for the primary mirror [9].\par
\smallskip
\noindent
The most recent UN/ESA Workshop on Basic Space Science: Scientific Exploration from Space, was hosted by the Institute of Astronomy and Space Sciences at Al al-Bayt University from 13 to 17 March 1999 in Mafraq, Jordan. The major result of the working group sessions was the urgent recommendation to make the small astronomical telescope facility (40-cm Maede telescope) on campus of Al al-Bayt University operational and to encourage the project of the construction of the  32-m Baquaa radio telescope at the University of Jordan, Amman [10].\par
\medskip
\noindent
{\bf 4. Third Cycle of Workshops: Networking Telescopes and Beyond}\par
\smallskip
\noindent
Based on the request from the United Nations, the Foreign Office of the Government of Germany, through the German Space Agency (DLR), made it possible to hold a UN/ESA Workshop on Basic Space Science at the Max-Planck-Institute for Radioastronomy (MPIfR), Bonn, Germany, in 1996, for the benefit of Europe. This workshop analyzed the results of all previous Workshops on Basic Space Science, particularly the follow-up projects that emanated from the second workshop cycle and charted the course to be followed in the future [6-10]. Additional to this objective, the workshop addressed scientific topics at the forefront of research in such diverse fields as photon, neutrino, gravitational waves, and cosmic rays astronomy, respectively. Taking into account that the past workshops did not lead yet to the establishment of an astronomical facility in African countries under consideration for such an effort, this workshop prepared the publication, on a regular basis,  of an urgently needed bilingual newsletter (African Skies/Cieux Africains) for the space science community in Africa, a  collaborative effort of astronomers from France and South Africa [11].\par
\smallskip
\noindent
The forthcoming Ninth UN/ESA Workshop on Basic Space Science: Satellites and Networks of Telescopes as Tools for Global Participation in the Studies of the Universe, will be held at Toulouse, France, in June 2000. The organizers of the series of workshops have agreed, based on observations and recommendations of the past workshops, that the agenda of this workshop will focus on the following topics:\par
\smallskip
\noindent
(i) Feasibility of the establishment of a World Space Observatory (WSO) [12].\par
\smallskip
\noindent
(ii) Network of Oriental Robotic Telescopes (NORT) [13].\par
\smallskip
\noindent
(iii) Networking of small astronomical telescopes to be preferentially utilized for observation of variable stars. The establishment of small astronomical telescope facilities with the sponsorship of Japan in Paraguay and the Philippines. Cooperation between small astronomical telescope facilities in terms of education and research programmes [14, 19] .\par
\smallskip
\noindent
(iv) Research with astronomical data bases [15] and the utilization of astrophysics data systems [16].\par
\medskip
\noindent
{\bf 5. Results That Supplemented the Workshop Series}\par
\smallskip
\noindent
1992 had been designated as International Space Year (ISY) by a wide variety of national and international space organizations, including the United Nations. To help generate interest and support for planetariums as centres of culture and education, the United Nations in cooperation with the International Planetarium Society, as part of its International Space Year activities, published a guidebook on the Planetarium: A Challenge for Educators [17]. Subsequently, this booklet was translated by national planetarium associations from English into Japanese, Slovakian, and Spanish, and is still available from the United Nations. \par
\smallskip
\noindent
In 1993, the European Space Agency, through the United Nations, donated 30 personal computer systems for use at universities and research institutions in Cuba, Ghana, Honduras, Nigeria, Peru, and Sri Lanka. \par
\smallskip
\noindent
In 1995, scientists from around the world gathered at the United Nations Headquarters in New York to discuss a broad range of scientific issues associated with near-Earth objects (NEOs). This gathering became known as the first United Nations International Conference on Near-Earth Objects [18]. Subsequently, the European Space Agency sponsored a study of a global network for research on near-Earth objects with the purpose to design and implement a worldwide information and data exchange centre called Spaceguard Central Node (SCN) in order to support follow-up activities after the detection of NEOs [19].\par
\medskip
\noindent
{\bf Acknowledgement}\par
\smallskip
\noindent
The author is deeply indebted to Dr. W. Wamsteker (European Space Agency) for his continual support and commitment in organizing the Workshops. The author would like to thank Prof. M. Kitamura (National Astronomical Observatory Tokyo), Dr. K.-U. Schrogl (German Space Agency Cologne), Dr. J. Andersen (International Astronomical Union Paris), and Prof. A.M. Mathai (McGill University Montreal) for their support of the Workshops.\par
\bigskip
\noindent
{\bf References}\par
\medskip
\noindent
Note: The author is writing in his personal capacity and the views expressed in this paper are those of the author and not necessarily those of the United Nations.\par
\medskip
\noindent
[1] For a comprehensive review of social and economic dimensions of science as a
collaborative effort see C.H. Lai (editor), Ideals and Realities: Selected Essays of Abdus Salam, 2nd edition (Singapore: World Scientific, 1987) and, focusing on the historic dimension of such endeavors, see [19]. \par
\smallskip
\noindent
[2] See J.N. Bahcall, The Decade of Discovery in Astronomy and Astrophysics (Washington, D.C.: National Academy Press, 1991). For the interplay on how technology of astronomical instruments, astrophysics, and mathematics produce the remarkable picture of the Universe, in the course of history of humankind, see R. Osserman, Poetry of the Universe: A Mathematical Exploration of the Cosmos (New York: Doubleday, 1995).\par
\smallskip
\noindent
[3] See J.R. Percy and A.H. Batten, "Chasing the dream", Mercury, 1995, 24(2):15-18. For an elaboration on the United Nations contributions see H.J. Haubold and W. Wamsteker, "Worldwide Development of Astronomy: The Story of a Decade of UN/ESA Workshops on Basic Space Science", Space Technology, 1998, 18(4-6):149-156, and H.J. Haubold, "UN/ESA Workshops on Basic Space Science: an initiative in the world-wide development of astronomy", Journal of Astronomical History and Heritage, 1998, 1(2):105-121.\par
\smallskip
\noindent
[4] Subsequently, these workshops were co-organized by the Austrian Space Agency (ASA), French Space Agency (CNES), German Space Agency (DLR), European Space Agency (ESA), International Astronomical Union (IAU),  International Centre for Theoretical Physics Trieste (ICTP), Institute of Space and Astronautical Science of Japan (ISAS), National Aeronautics and Space Administration of the United States (NASA), The Planetary Society (TPS), and the United Nations (UN).\par
\smallskip
\noindent
[5] A month-to-month update on results and new developments related to the UN/ESA Workshops on Basic Space Science is made available at the Workshop's World-Wide-Web site at http://www.seas.columbia.edu/$\sim$ah297/un-esa/. The Proceedings of the workshops were published in: (I) Conference Proceedings of the American Institute of Physics Vol. 245, American Institute of Physics, New York, 1992, pp. 350; (II) Earth, Moon, and Planets 63, No. 2 (1993)93-170; (III) Astrophysics and Space Science 214, Nos. 1-2 (1994)1-260; (IV) Conference Proceedings of the American Institute of Physics Vol. 320, American Institute
of Physics, New York, 1994, 320pp.; (V) Earth, Moon, and Planets 70, Nos. 1-3 (1995)1-233; (VI) Astrophysics and Space Science 228, Nos. 1-2 (1995)1-405; and (VII) Astrophysics and Space Science 258, Nos. 1-2 (1998)1-394.\par
\smallskip
\noindent
[6] For a detailed report on the development of the GEM project, its scientific results, impacts on university education and research in Colombia, and references to the literature see S. Torres, ``The UN/ESA Workshop on Basic Space Science in Colombia, 1992: What has been achieved since then?'', COSPAR Information Bulletin, 1999, No. 144, pp. 13-15. The World-Wide-Web site of GEM can be accessed at http://aether.lbl.gov/www/projects/GEM. \par
\smallskip
\noindent
[7] S. Gunasekara and P. de Alwis, "The astronomy promotional programme at ACCIMT", in Conference on Space Sciences and Technology Application for National Development: Proceedings, held at Colombo, Sri Lanka, 21-22 January 1999, Ministry of Science and Technology of Sri Lanka, pp. 143-146. The World-Wide-Web site of ACCIMT at http://www.slt.lk/accimt/page5.html is gradually incorporating results obtained with the telescope facility. See also the papers of Kitamura and Kogure, respectively, in [14].\par
\smallskip
\noindent
[8] The Observatory and its educational and scientific activities is part of the
World-Wide-Web site at http://www.unah.hondunet.net/unah.html. A recent  photograph of the Observatory building is available at\\  
http://www.laprensahn.com/natarc/9812/n23001.htm. \par
\smallskip
\noindent
[9] S.M. Hasan, ``Upgrading the 1.9-m Kottamia telescope'', African Skies, 1998, No. 2, pp. 16-17.\par
\smallskip
\noindent
[10] For all workshops, United Nations Reports on the organization of the respective Workshop have been published as UN General Assembly documents, see Report on the Eighth United Nations/European Space Agency Workshop on Basic Space Science: Scientific Exploration from Space, hosted by the Institute of Astronomy and Space Sciences at Al al-Bayt University on behalf of the Government of Jordan, A/AC.105/723, 18 May 1999, 8pp. A special World-Wide-Web site was developed for this workshop at http://www.planetary.org/news/Events/unispace.html.\par
\smallskip
\noindent
[11] The World-Wide-Web site of the Working Group for Space Science in Africa is
http://da.saao.ac.za:80/$\sim$wgssa/. In this connection, see, for a detailed review of the workshop in Nigeria, held in 1994, L.I. Onuora, ``The UN/ESA Workshop on Basic Space Science in Nigeria: Looking back'', COSPAR Information Bulletin, 1999, No. 144, pp. 15-16.\par
\smallskip
\noindent
[12] W. Wamsteker and R. Gonzales Riestra (editors), Ultraviolet astrophysics beyond the IUE final archive: Proceedings of the conference, held at Sevilla, Spain, 11-14 November 1997, European Space Agency SP-413, pp. 849-855. H. Gavaghan, ``U.N. plans its future in space'', Science, 1999, 285, p. 819. See also Report on the Eighth United Nations/European Space Agency Workshop on Basic Space Science: Scientific Exploration from Space, hosted by the Institute of Astronomy and Space Sciences at Al al-Bayt University on behalf of the Government of Jordan, A/AC.105/723, 18 May 1999, 8pp.\par
\smallskip
\noindent
[13] F.R. Querci and M. Querci, ``The network of oriental robotic telescopes'', African Skies, 1998, No. 2, pp. 18-21.\par
\smallskip
\noindent
[14] H. Gavaghan, ``U.N. plans its future in space'', Science, 1999, 285, p. 819. M. Kitamura, ``Provision of astronomical instruments to developing countries by Japanese ODA with emphasis on research observations by the donated 45-cm reflectors in Asia'', in Conference on Space Sciences and Technology Application for National Development: Proceedings, held at Colombo, Sri Lanka, 21-22 January 1999, Ministry of Science and Technology of Sri Lanka, pp. 147-152. T. Kogure, "Stellar activity and needs for multi-site observations", in Conference on Space Sciences and Technology Application for National Development: Proceedings, held at Colombo, Sri Lanka, 21-22 January 1999, Ministry of Science and Technology of Sri Lanka, pp. 124-131.\par
\smallskip
\noindent
[15] E.g., IUE Newly Extracted Spectra (INES), World-Wide-Web site at
http://ines.vilspa.esa.es, which is a complete astronomical archive and data distribution system, representing the final activity of ESA in the context of the International Ultraviolet Explorer project.\par
\smallskip
\noindent
[16] E.g., the NASA Astrophysics Data System (ADS), World-Wide-Web site at
http://adswww.harvard.edu, whose main resource is an abstract service which includes four sets of abstracts: (i) astronomy and astrophysics, (ii) instrumentation, (iii) physics and geophysics, and (iv) Los Alamos preprint server.\par
\smallskip
\noindent
[17] See World-Wide-Web site at
http://www.seas.columbia.edu/$\sim$ah297/un-esa/planetarium.html.\par
\smallskip
\noindent
[18] See World-Wide-Web site at 
http://www.seas.columbia.edu/$\sim$ah297/un-esa/neo.html.\par
\smallskip
\noindent
[19] See World-Wide-Web site at
http://spaceguard.ias.rm.cnr.it\par
\smallskip
\noindent
[20] L. Pyenson and S. Sheets-Pyenson, Servants of Nature: A History of Scientific Institutions, Enterprises, and Sensibilities, W.W. Norton \& Company, New York, 1999, pp. XV+496.\par
\smallskip
\noindent
[21] The report on this UNISPACE III Conference is available electronically at
http://www.un.or.at/OOSA; as part of the Technical Forum of UNISPACE III, comprising 38 scientific activities, an IAU/COSPAR/UN Special Workshop on Education in Astronomy and Basic Space Science was held leading to conclusions and proposals contained in UN Document\\ A/CONF.184/C.1/L.8.
\end{document}